\documentclass[aps, twocolumn, prb]{revtex4}
\usepackage{amssymb}
\usepackage{amsmath}
\usepackage[font=footnotesize,labelfont=bf,justification=raggedright,figurename=Figure]{caption}
\usepackage{float}
\usepackage{bbm}
\usepackage{graphicx}
\usepackage{verbatim}
\usepackage[normalem]{ulem}
\usepackage{graphicx,amsmath}

\newcommand{\beq}{\begin{equation}}
\newcommand{\eeq}{\end{equation}}
\newcommand{\beqarray}{\begin{eqnarray}}
\newcommand{\eeqarray}{\end{eqnarray}}
\usepackage{hyperref}
\usepackage{xcolor}
\hypersetup{
    colorlinks,
    citecolor=blue,
    filecolor=blue,
    linkcolor=blue,
    urlcolor=blue,
    pdfstartview=FitH
}

\def\npsection#1{{\par{\vskip 5pt}\noindent\bf #1}}

\def\papertitle{Majorana fermions in ferromagnetic chains on the surface of bulk spin-orbit coupled $s$-wave superconductors} 
\begin{document}

\title{\papertitle}
\author{Hoi-Yin Hui$^{1}$, P. M. R. Brydon$^{1}$, Jay D. Sau$^{1}$, S. Tewari$^{1,2}$ \& S. Das Sarma$^{1}$} 
\affiliation{$^{1}$Department of Physics, Condensed Matter Theory Center and Joint Quantum Institute, University of Maryland, College Park, MD 20742, USA}
\affiliation{$^{2}$Current address: Department of Physics and Astronomy, Clemson University, Clemson, SC 29634, USA}

\maketitle

{\bf
Majorana fermion (MF) excitations in solid state system have
non-Abelian statistics which is essential for topological quantum
computation. Previous proposals to realize MF, however,
generally requires fine-tuning of parameters.
Here we explore a platform which avoids the fine-tuning problem,
namely a ferromagnetic chain deposited 
on the surface of a spin-orbit coupled $s$-wave superconductor. We
show that it generically supports zero-energy topological MF
excitations near the two ends of the chain with minimal
fine-tuning. Depending on the strength of the ferromagnetic moment in
the chain, the number of MFs at each end, $n$, can be either one or
two, and should be revealed by a robust zero-bias peak (ZBP) of height
$2ne^2/h$ in scanning  tunneling microscopy (STM) measurements which
would show strong (weak) signals at the ends (middle) of the
chain. The role of an approximate chiral symmetry which gives an integer
topological invariant to the system is discussed.
}

\maketitle

Majorana fermions (MFs), which are also their anti-fermions by
definition, were originally introduced almost 80 years ago in the
context of understanding neutrinos. As real solutions of the Dirac
equation they are therefore self-conjugate, thus representing both
particles and anti-particles by the same real wavefunction. Whether
neutrinos are MFs or not is still an open question being vigorously
investigated in high-energy physics. A completely novel incarnation of
MFs appeared more recently
\cite{Kitaev2001Unpaired,Nayak2008Non-Abelian,Read2000Paired} in
condensed matter physics as zero-energy neutral bound states in the
subgap excitation spectrum of topological superconductors (TS). The
exact particle-hole symmetry characteristic of the excitation spectrum
in superconductors usually sharply distinguishes states as either
electron-like or hole-like in accordance with whether their energy is
positive or negative, respectively. In $s$-wave superconductors the
fermion-doubling theorem prevents the appearance of any zero-energy
subgap excitations, and so MFs can only appear in effectively spinless
$p$-wave superconductors, which are the canonical example of TS with a
bulk gap and topologically protected gapless edge states. Such
Majorana bound states (i.e. MFs in TS) in low-dimensional condensed
matter systems obey anyonic non-Abelian braiding statistics and are in
general anyons, not ordinary fermions, which makes them ideal for
fault-tolerant quantum computation
\cite{Kitaev2001Unpaired,Nayak2008Non-Abelian,Read2000Paired}. These
MFs in TS systems typically arise at defect sites (e.g. vortex cores,
interfaces and edges) as localized excitations, and are topologically
protected against local perturbations by the bulk superconducting
gap. Although spinless $p$-wave superconductors do not seem to exist
in nature, there have been many recent proposals
\cite{Fu2008Superconducting,Zhang2008pxipy,Sato2009Non-Abelian,Lutchyn2010Majorana,Oreg2010Helical,Chung2011Topological,Duckheim2011Andreev,Choy2011Majorana,Nadj-Perge2013Proposal}
for artificially creating two- and one-dimensional systems which
behave effectively as spinless $p$-wave
superconductors which support MFs at their boundaries. There are
even experimental claims of the possible observation
\cite{Mourik2012Signatures,Deng2012Anomalous,Rokhinson2012fractional,Das2012Zero-bias,Churchill2013Superconductor-nanowire,Finck2013Anomalous}
of MFs in spin-orbit-coupled semiconductor-superconductor
heterostructures following theoretical proposals
\cite{Lutchyn2010Majorana,Oreg2010Helical}, but the situation is not
definitively conclusive. Given the great fundamental and practical
significance of MFs, it is desirable to have platforms where MFs could
easily emerge for experimental 
observation and investigation. 

In this paper we study theoretically a relatively simple
scheme for realizing MFs 
in a condensed matter setting, involving ferromagnetic (presumably
metallic, e.g. Fe or Co or Ni) chains placed on the surface of
standard superconductors (e.g. Pb or Nb or Al) in an STM-type
measurement set up \cite{Menzel2012Information}. The advantage of this
scheme is that MFs in the proposed system generically occur in a
wide range of parameter space, thus requiring little fine-tuning of system
parameters (e.g. tuning the applied magnetic field appropriately as in the
semiconductor heterostructure scheme
\cite{Lutchyn2010Majorana,Oreg2010Helical}) although  
for some values of the parameters more than one MFs are spatially
superimposed on each other. Consequently, a zero-bias tunneling peak
(ZBP), which is a hallmark of the zero-energy MFs 
\cite{Law2009Majorana,Flensberg2010Tunneling}, is robust and generic
in the proposed system. 
Our work significantly extends the robustness of previous
proposals to realize MFs in superconductor-ferromagnet heterostructures
\cite{Chung2011Topological,Duckheim2011Andreev}, thus making such
devices an attractive alternative to spin-orbit-coupled
semiconductor-superconductor platforms.

\begin{figure*}
\includegraphics[width=0.6\textwidth]{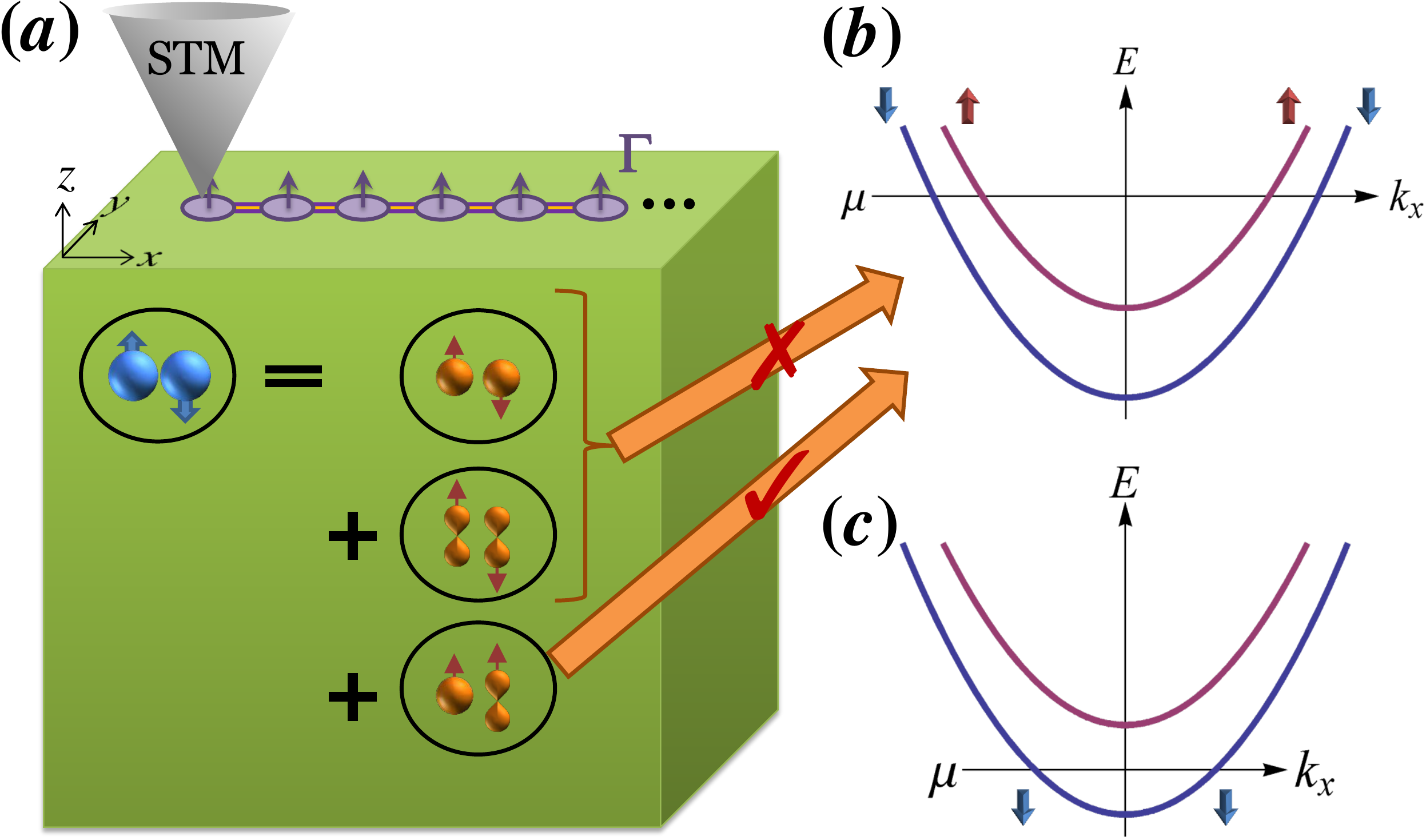} 
\caption{\label{fig:device} (a) Schematic diagram of our device. A
  ferromagnetic (FM) chain is placed on the surface of an $s$-wave
  superconductor, in which 
  strong spin-orbit coupling and mixing of orbitals of opposite parity
  produce a pairing state with intra-orbital spin-singlet Cooper pairs
  and inter-orbital spin-triplet pairs. Tunneling of these pairs into
  the chain generates effective spin-singlet and spin-triplet pairing
  potentials, respectively, as shown in (b). For a FM chain 
  with spin-splitting that exceeds the spin-singlet
  pairing potential, only the induced triplet pairing potential can 
  gap the spectrum. In this case the system is in a topologically
  nontrivial state 
  characterized by two unhybridized Majorana fermions at each end,
  which can be imaged by scanning tunneling microscopy (STM). When
  the FM chain is in the half-metal regime as shown in panel
  (c), however, only a non-Abelian single Majorana fermion is realized at each
  end. If the spin-splitting of the chain states is much smaller than
  their bandwidth, however, the situation (b) dominates the parameter
  space.}
\end{figure*}

In a conventional $Z_2$ topological superconductor
\cite{Schnyder2008Classification,Schnyder2009Classification,Kitaev2009Periodic,Ryu2010Topological}, 
e.g., the spin-orbit-coupled semiconductor-superconductor heterostructure
\cite{Lutchyn2010Majorana,Oreg2010Helical}, a pair of MFs spatially superimposed
on each other mix and split to finite energies, thus essentially becoming low-energy fermionic subgap states \cite{DasSarma2012Splitting}. In this class of TS
systems, therefore, the number of MFs ($n$) at any point in space can
be either zero or one. This topological restriction on $n$ results in
a greatly reduced parameter space in which to look for experimental
signatures of MFs. In the semiconductor-superconductor nanowire
heterostructure, for example, ZBPs are expected in the presence of an
externally applied magnetic field 
-- the so-called semiconductor Majorana wire -- only
when the number of semiconductor bands crossing the Fermi energy is
odd \cite{Lutchyn2011Search,Potter2010Multichannel}, a condition
difficult to control experimentally. Similarly, proposals for
realizing a Majorana fermion in ferromagnet-superconductor
heterostructures \cite{Chung2011Topological,Duckheim2011Andreev} have
the stringent requirement that only one of the
spin-split bands in the ferromagnet has a Fermi surface.
In the experimental system we explore in this paper, strictly
speaking, there is no 
topological restriction on the number of MFs that can be localized at
a given point in space. As we show below, the absence of a restriction
on $n$, resulting from a topological chiral symmetry
\cite{Schnyder2008Classification,Schnyder2009Classification,Kitaev2009Periodic,Ryu2010Topological},
results in a greatly enhanced parameter space in which MFs are realized.
We emphasize, however, that only when 
$n$ is odd does the Majorana multiplet follow non-Abelian braiding statistics
although a robust ZBP in STM experiments should occur 
generically for any value of $n$. Of course, for the purpose of 
establishing topologically protected degenerate states that may be 
used to establish non-Abelian braiding \cite{Read2000Paired}, it is
necessary for the Majorana to  
be non-degenerate i.e. $n=1$, and therefore the generic ZBP signature
here cannot necessarily be identified with a non-Abelian Majorana
``particle''. Our conceptual new finding that robust Majorana fermions
may reside generically (i.e. with no fine-tuning) in
superconductor-ferromagnet heterostructures, protected by a hitherto
undiscovered chiral symmetry, is the important new result presented in
this theoretical work.

\npsection{Results}

\begin{figure}
\begin{centering}
\includegraphics[width=0.6\columnwidth]{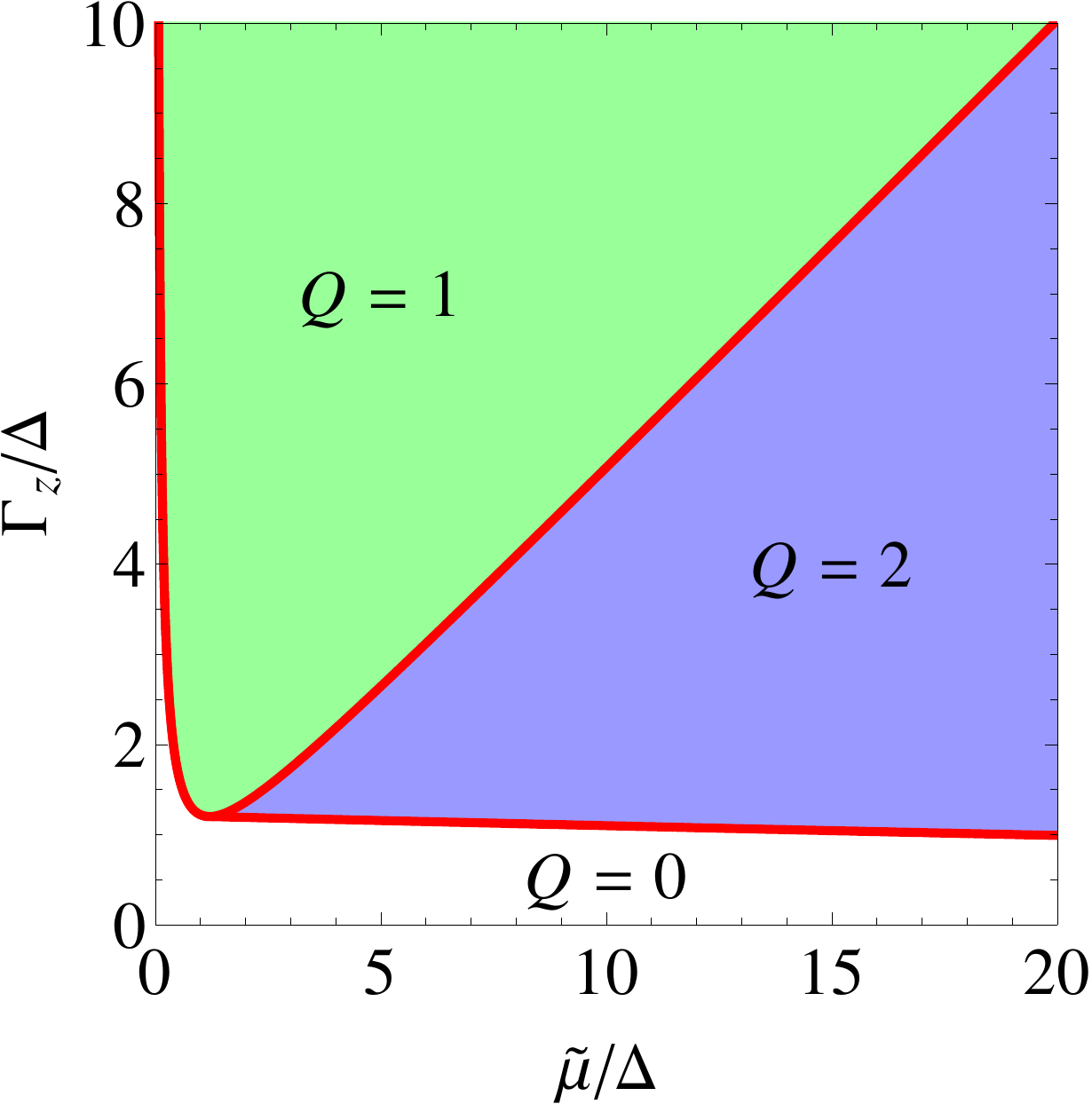}
\par\end{centering}
\caption{ {\bf Topological phase diagram of the chain.} The BDI topological index $Q$ is defined in equation (\ref{eq:Q})
  calculated for $H_{{\rm chain}}^{{\rm eff}}$ as a function of the
  Zeeman splitting $\Gamma_z$ and the chemical potential. The Green
  region (roughly $\Gamma_z>\tilde{\mu}/2$) has $Q=1$ while the blue
  region (roughly $\Gamma_z<\tilde{\mu}/2$) has $Q=2$, indicating the
  existence of one and two Majorana fermions at each end of the chain,
  respectively.\label{fig:Phase}}
\end{figure}

\begin{figure}
\begin{centering}
\includegraphics[width=0.8\columnwidth]{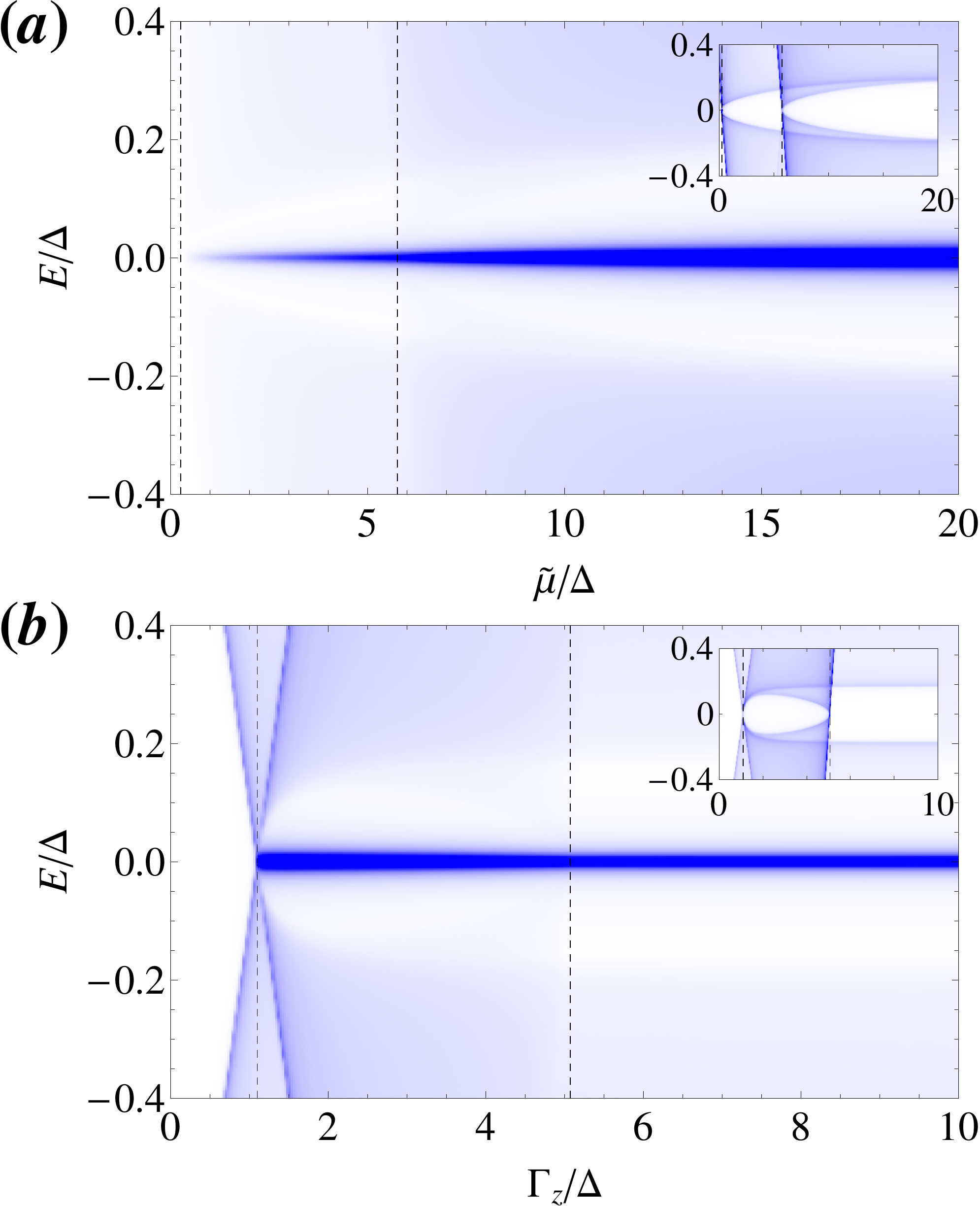}
\par\end{centering}
\caption{(a) The local density of states (LDOS) at the ends of the semi-infinite
  ferromagnetic (FM) chain as a function of the chemical potential
  $\tilde{\mu}$ from the bottom of the bands, for $\Gamma_z = 3\Delta$. The
  LDOS has a strong zero-energy peak that for roughly $\tilde{\mu} < 2 \Gamma_z$
  indicates a single Majorana fermion (MF) from the chain ends, while for  
 $\tilde{\mu}>2\Gamma_z$ the zero-bias peak implies a pair of MFs
  localized at each ends protected by chiral symmetry.  (b) the LDOS 
at the chain ends as a function of the Zeeman splitting for
$\tilde{\mu}=10\Delta$. For roughly $\Gamma_z < \tilde{\mu}/2$ ($\Gamma_z >
\tilde{\mu}/2$) the zero-energy peak in LDOS signifies two (one) MFs at
each end that can be accessed in scanning tunneling microscopy
experiments. The insets shows the LDOS at the middle of the chain,
which has a spectral gap in the topological regions. We indicate the
transitions between the different topological sectors by the vertical
dashed lines, and use arbitrary units for the LDOS in these
plots. \label{fig:ldosMuB}}  
\end{figure}

\npsection{Ferromagnetic chain on a spin-orbit coupled superconductor.}
A ferromagnetic (FM) chain (e.g. Fe), which is a single atom in
width (although a few atoms should work too), is placed on the surface of a bulk $s$-wave 
superconductor, as shown schematically in Fig.~\ref{fig:device}.  
We emphasize that, in contrast to arrays of magnetic atoms on the
surface of a superconductor, the FM chain is expected to
have a bandwidth that is orders of  
magnitude larger than the superconducting pairing potential.
We ignore the spin-orbit coupling within the
FM chain (which plays no role in the scheme whether it exists or
not, in sharp contrast to the semiconductor heterostructure-based
Majorana schemes), but instead account for the existence of strong inversion-symmetric
spin-orbit coupling in the bulk of the host superconductor. By
integrating out the bulk superconductor we show that the effective
Hamiltonian of the FM chain [equation (\ref{eq:effwire2}) below] is in the chiral BDI
class with an integer invariant, allowing an integer number $n$ of 
MFs localized at the chain ends. If the FM chain has only one pair of
spin-split sub-bands, $n$ can be equal to zero, one, or two, but for
\textit{any} non-zero $n$ (a  condition that is realized in most of
the parameter space (Fig.~\ref{fig:Phase})) STM measurements at the
chain ends should reveal a pronounced ZBP. The ZBP is in fact generic
in our model, occurring in a wide region of the
experimentally-accessible parameter
space as shown in Fig.~\ref{fig:Phase} and \ref{fig:ldosMuB}. No such peak is
expected from the regions of the chain away from the ends where the MFs are localized. In practice
the effective chiral symmetry in the FM chain should only be
approximate, resulting in a finite energy width of the ZBPs for
$n>1$.

\npsection{Effective Hamiltonian of the chain.}
The superconductor used in our device must satisfy two key
conditions: (i) there is strong spin-orbit coupling
\cite{Kim2014Helical}, although 
inversion symmetry is not necessarily broken in the bulk, and (ii)
orbitals of different 
parity both make a significant contribution to the states near the
Fermi surface. 
The requirement that the orbitals have opposite parity can be
  relaxed, but this condition makes the following argument more transparent.
The first condition implies that spin is not a good quantum number in
the superconductor, but the presence of time-reversal
(${\cal T}$)  and inversion (${\cal I}$) symmetry means that the
doubly-degenerate eigenstates at each momentum ${\bf k}$ can be labeled by
a pseudospin index 
$\varsigma = \pm $, such that ${\cal T}|{\bf 
  k},\varsigma \rangle = \varsigma |-{\bf k},-\varsigma\rangle$ and
${\cal I}|{\bf k},\varsigma\rangle = |-{\bf
  k},\varsigma\rangle$. A conventional $s$-wave superconducting gap
then corresponds to a pseudospin-singlet pairing state.  
To satisfy the second condition, we assume that the states near
the Fermi surface are composed from two orbitals, say $s$ and $p$,
which are symmetric and antisymmetric under inversion,
respectively. The general form of the pseudospin state is then 
\begin{equation}
|{\bf k},\varsigma\rangle = \sum_{\sigma=\uparrow,\downarrow}\left\{B^s_{\varsigma,\sigma}({\bf k})|s,{\bf
  k},\sigma\rangle + B^p_{\varsigma,\sigma}({\bf k})|p,{\bf
  k},\sigma\rangle\right\}\,. \label{eq:pseudospin}
\end{equation}
Due to the different parities of the two orbitals, the
coefficients of the $s$ and $p$ states are even and odd in ${\bf k}$,
respectively. Expressed in the orbital basis
using equation (\ref{eq:pseudospin}), one generally finds that the
pseudospin-singlet pairing potential includes both intra-orbital
spin-singlet and inter-orbital spin-triplet terms. The latter play a
critical role in generating the topological state in 
the magnetic chain. 

The tunneling between the magnetic chain and the superconductor is
assumed to be local and independent of spin, and is therefore most
transparently formulated in terms of tunneling between the chain atoms
and the adjacent orbitals of the superconductor. We assume the form
\begin{equation}
H_{\text{tun}}=\sum_{{\bf r}\in\text{chain}}\sum_{\sigma}\left\{ f_{{\bf r},\sigma}^{\dagger}[t_{s}s_{{\bf r},\sigma}+t_{p}p_{{\bf r},\sigma}]+\text{H.c.}\right\}
\end{equation}
 where $t_{s}$ and $t_{p}$ are the tunneling matrix elements for
the two orbitals, assumed real, and $f_{{\bf r},\sigma}$, $s_{{\bf
    r},\sigma}$ and $p_{{\bf r},\sigma}$ are the annihilation
operators for the site 
${\bf r}$ in the chain and in the superconductor's $s$ and $p$
orbitals, respectively. The tunneling Hamiltonian implicitly
  accounts for the surface inversion-symmetry breaking: if the
  odd-parity orbital is odd with respect to mirror reflection in the
  surface plane, then tunneling from the chain sites into both the even-
  and odd-parity orbitals of the underlying superconductor can have a
  local component (see~Fig.~(\ref{fig:device})).
Since we are interested in the physics
of the chain, our strategy is now to ``trace out'' the superconductor
from the description of the problem. After standard
manipulations as detailed in the Methods section, we obtain the
self-energy correction for the chain
\begin{equation}
\Sigma(x,x';\omega) = {\bf T} G_{{\rm orb}}(x,x';\omega) {\bf T}^\dagger, \label{eq:selfenergy}
\end{equation}
where the matrix ${\bf T}$ describes
the tunneling between the chain and the superconductor, while
$G_{{\rm orb}}(x,x';\omega)$ is the Green's function of the
superconductor expressed in the orbital-spin basis, and is related
to the pseudospin Green's function by
\begin{equation}
G_{\text{orb}}({\bf k},\omega)=\left(\begin{array}{cc}
\check{B}_{{\bf k}}^T & 0\\
0 & \check{B}_{{\bf k}}^T
\end{array}\right)G_{\text{pseudo}}({\bf k},\omega)\left(\begin{array}{cc}
\check{B}_{{\bf k}}^\ast & 0\\
0 & \check{B}_{{\bf k}}^\ast
\end{array}\right)\,,
\end{equation}
where $\check{B}_{{\bf k}}=(\hat{B}^{s}({\bf k}),\hat{B}^{p}({\bf k}))$ is the matrix of coefficients in equation (\ref{eq:pseudospin}).
The self-energy contains a complicated proximity-induced pairing
term. Crucially, we find a $p$-wave spin-triplet pairing due 
to the tunneling of inter-orbital Cooper pairs from the superconductor.
We generally expect that there will be triplet Cooper pairs with spin
parallel to the magnetization, and so a gap appears in the spin-split
states 
of the FM chain, see~Fig.~\ref{fig:device}(b). By contrast,
the spin-singlet pairing due to tunneling of intra-orbital Cooper pairs
is unable to overcome the large exchange splitting. The proximity
effect also renormalizes the bare dispersion and produces a spin-orbit
coupling, but these effects are small and will be ignored.

The general expressions for the self-energy is unenlightening and
presented in the Methods section. To make progress
we write the general forms of the coefficient matrices
$\hat{B}^{s}({\bf
  k})=\alpha^{s}_{\bf k}\hat{1}+i\pmb{\beta}^{s}_{\bf k}\cdot{\hat{\pmb{\sigma}}}$ 
and $\hat{B}^{t}({\bf k})=\alpha^{p}_{\bf k}{\bf e}_{\bf
  k}\cdot{\hat{\pmb{\sigma}}}+\pmb{\beta}^{p}_{\bf k}\cdot({\bf e}_{\bf
  k}\times{\hat{\pmb{\sigma}}}) + i\pmb{\gamma}^p_{\bf k}\cdot{\bf e}_{\bf
  k}$, where ${\bf e}_{\bf k}$ is the unit vector in the direction of
${\bf k}$, and the
coefficients are all real and even functions of the momentum.
The presence of mirror symmetries along three directions in the bulk, and two on its surface,
further constrains the possible forms for $B^s$ and $B^p$. To be
concrete, we choose $\alpha^s_{\bf k}=\alpha^p$ and 
$\pmb{\beta}^{p}_{\bf k}=\beta^{p}{\bf e}_{z}$ which lead to terms consistent with the mirror symmetry requirements. We
further assume that the 
coupling between the chain and the superconductor is small compared with
the chain's bandwidth so that the induced gap is much less than the
other energy scales 
of the system. We then obtain an effective Hamiltonian by adding the
self energy [equation (\ref{eq:selfenergy})] evaluated at $\omega=0$ to the
bare chain Hamiltonian.
Neglecting corrections beyond nearest-neighbor pairing,
we obtain the effective Hamiltonian of the chain
\begin{eqnarray}
H_{{\rm chain}}^{{\rm eff}}\left(k_{x}\right) & = & \left(-2t\cos k_{x}-\mu\right)\hat{\tau}_{z}+{\bf \Gamma}\cdot\hat{\pmb{\sigma}}\nonumber \\
 &  & +\left(\Delta+\tilde{\Delta}\cos k_{x}\right)\hat{\tau}_{x} +\tilde{\Delta}^{(t)}\sin k_{x}\hat{\sigma}_{y}\hat{\tau}_{x}\,,
 \label{eq:effwire2}
\end{eqnarray}
where $\hat{\sigma}_\mu$ and $\hat{\tau}_\mu$ are the Pauli matrices in
spin and Nambu space, respectively.
The first line of the Hamiltonian describes the bare FM
chain with direct inter-atom hopping $t$, chemical potential $\mu$, and
a Zeeman splitting ${\bf \Gamma}\cdot\hat{\pmb{\sigma}}$ due to
ferromagnetism which is comparable to the Fermi energy in the wire.
 The last line gives the induced superconducting gaps 
with both singlet ($\Delta$ and $\tilde{\Delta}$) and triplet
($\tilde{\Delta}^{(t)}$) pairing potentials. The latter corresponds to
a state where the triplet pairs have vanishing spin component along the
$y$-axis which can gap the spin-split bands as long as ${\bf \Gamma}$ has a
component in the $x$-$z$ plane. 

 The key experimentally-relevant quantity is the
local density of states (LDOS), which can be directly measured using
STM. The LDOS at position $x$ is
defined as
\begin{equation}
\nu\left(x,\omega\right)=\frac{-1}{2\pi}{\rm ImTr}\left[\omega+i\delta-H_{{\rm chain}}^{{\rm eff}}\left(x,x\right)\right]^{-1}\left(1+\hat{\tau}_{z}\right)\,.\label{eq:DOS}
\end{equation}
Throughout this paper we fix $t=10\Delta$ and
$\tilde{\Delta}=\tilde{\Delta}^{(t)}=0.2\Delta$, and study how the
topology of the system varies as a function of $\mu$ and
$\pmb{\Gamma}$. We emphasize that our results are generic and
qualitatively 
independent of the precise choice of these parameters.

\npsection{Topological properties of the chain.}
For obtaining the
topological classification of $H_{{\rm chain}}^{{\rm eff}}$
we note that it satisfies the particle-hole symmetry $\left\{ H_{{\rm
    chain}}^{{\rm eff}},{\hat\Xi}\right\} =0$, 
where ${\hat\Xi}=\sigma_{y}\tau_{y}K$ and $K$ is the complex-conjugate
operator. If we further assume that the $y$ component of ${\bf \Gamma}$
is zero, $H_{{\rm chain}}^{{\rm eff}}$ is real and hence it has the
chiral symmetry $\left\{ H_{{\rm chain}}^{{\rm eff}},\mathcal{\hat C}\right\} =0$
where $\mathcal{\hat C}=\sigma_{y}\tau_{y}$. In this case, $H_{{\rm chain}}^{{\rm eff}}$
belongs to the BDI topological class and is endowed with a topological
index $Q$ equal to the number of zero-energy MF modes ($n=Q$) localized at its
ends. To compute $Q$, we first
rotate $H_{{\rm chain}}^{{\rm eff}}$ 
to the basis in which ${\cal \hat C}$ is diagonal, by
${\hat U}=e^{-i\frac{\pi}{4}\tau_{x}\sigma_{y}}$, 
such that
\begin{equation}
\tilde{H}_{{\rm chain}}^{{\rm eff}}  ={\hat U}H_{{\rm chain}}^{{\rm eff}}{\hat U}^{\dagger} =\left(\begin{array}{cc}
0 & A_{k_{x}}\\
A_{-k_{x}}^{T} & 0
\end{array}\right),
\end{equation}
whence $Q$ is computed by
\begin{equation}
Q=\frac{1}{\pi}\int_{0}^{\pi}dk_{x}\frac{d\arg\left(\det A_{k_{x}}\right)}{dk_{x}}.
\label{eq:Q}
\end{equation}

In Fig.~\ref{fig:Phase} we plot $Q$ against $\tilde{\mu}$ and $\Gamma_{z}$,
where ${\bf \Gamma}$ is taken as ${\bf \Gamma}=\Gamma_{z}{\bf e}_z$ such that the
chiral symmetry is respected (the chiral symmetry is respected as long
as ${\bf \Gamma}$ is in the ($x$-$z$) plane), and $\tilde{\mu}=(\mu+2t+\Gamma_z)$
is defined from the bottom of the spin-split sub-bands. Note that for approximately
$\Delta<\Gamma_{z}<\tilde{\mu}/2$, 
we have $Q=2$ while for  $\Delta,\tilde{\mu}/2<\Gamma_z$
we have $Q=1$, indicating, in both cases, the existence of MFs at the
chain ends. This can be understood in the following way: in the large
Zeeman spin-splitting (``half-metal'')
limit, the effects of the singlet pairing terms $\Delta$ and
$\tilde{\Delta}$ on the Bogoliubov-de Gennes spectrum are suppressed
due to a large   
Fermi momenta mismatch between the two spin species. Then, with the
triplet pairing $\tilde{\Delta}^{(t)}$, the system becomes effectively an
equal-spin-pairing triplet superconductor with non-zero
$\Delta_{\uparrow\uparrow}$ and $\Delta_{\downarrow\downarrow}$, which can
be viewed as two copies of the Kitaev $p$-wave chain spatially
superimposed on each other \cite{Dumitrescu2013Topological}. If $\tilde{\mu}$ is
such that both spin channels are occupied, we get $Q=2$ (two MFs at
each end of the chain), while if $\tilde{\mu}$ is such that only 
one channel is occupied, we get $Q=1$ (an MF at each end). Since
most itinerant ferromagnets are not half metals, and the induced
singlet gap $\Delta$ is likely much smaller than $\tilde{\mu}$, we expect
that the $Q=2$ phase is of greatest practical relevance.

In Fig.~\ref{fig:ldosMuB}(a,b) we plot
the LDOS at the ends of the semi-infinite FM chain as a function of
$\tilde{\mu}$ and $\Gamma_z$, respectively. 
It can be seen that the LDOS generically has a pronounced zero-energy
peak that can be accessed in STM measurements near the
chain ends. For  
the Zeeman splitting $\tilde{\mu}>2\Gamma_z$ the ZBP is due to a pair of
MFs localized at the same ends and protected from splitting by the
topological chiral symmetry. For $\tilde{\mu}<2\Gamma_z$ the zero-energy
peak implies a single MF that should follow non-Abelian braiding
statistics. No such zero-energy peak is observed in LDOS calculated
for the middle of the chain, although the superconducting gap in the chain
closes at the 
topological transitions at which the integer $Q$ (and thus the number
of MFs at the chain ends) changes, see the inset of Fig.~\ref{fig:ldosMuB}(a,b).

\npsection{Discussion}

\begin{figure}
\begin{centering}
\includegraphics[width=0.8\columnwidth]{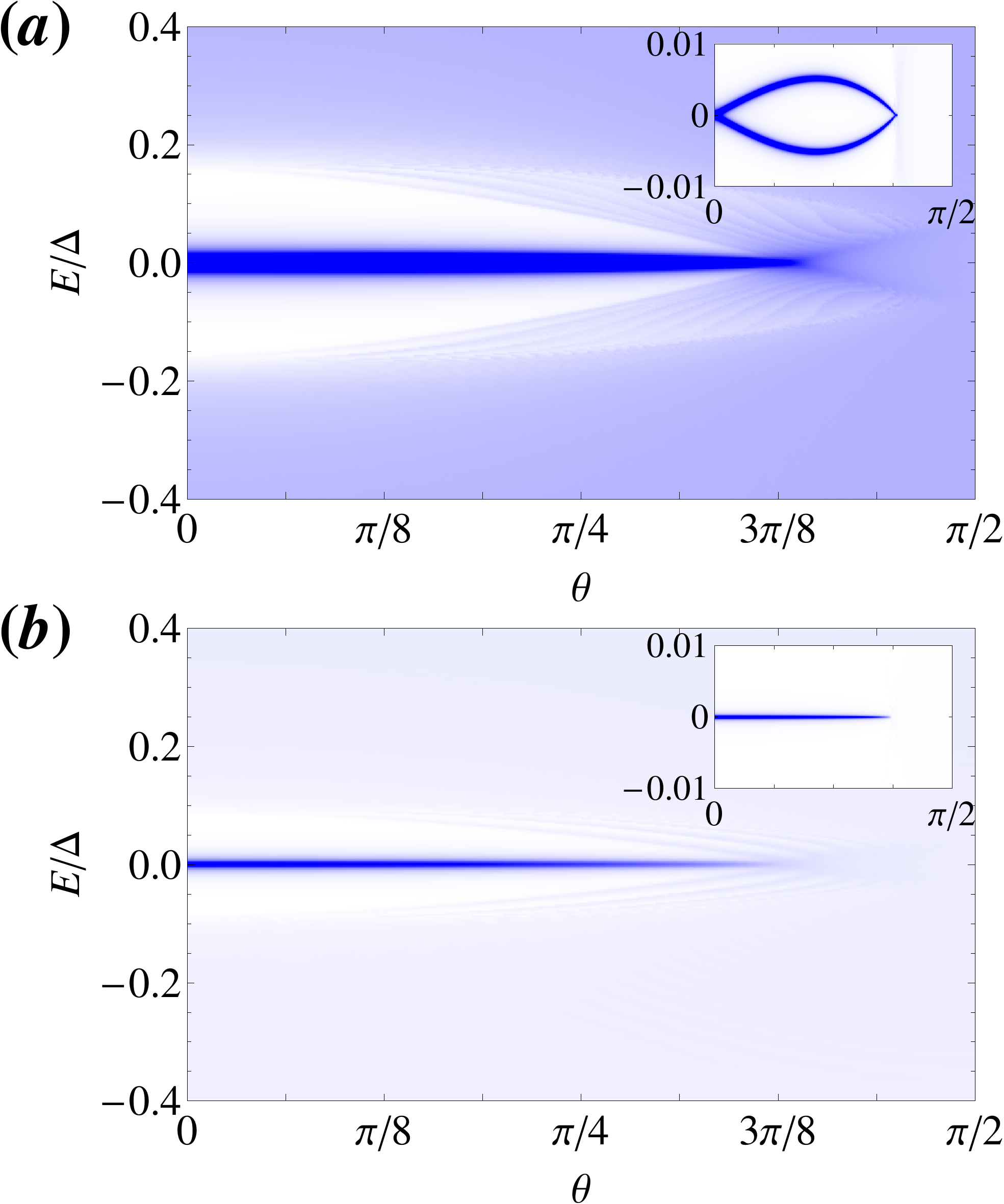}
\par\end{centering}
\caption{(a) The local density of states (LDOS) at one end of the semi-infinite
  ferromagnetic (FM) chain as a function of $\theta$ in the phase $Q=2$, where $\tilde{\mu}=15\Delta$ and ${\bf \Gamma}=3\Delta\left(\sin\theta{\bf e}_y+\cos\theta{\bf e}_z\right)$.
  Since the $y$-component of ${\bf \Gamma}$ breaks chiral symmetry, the pair of MFs at each end mix and split for finite $\theta$, but
 the splitting is small and visible only on a small energy scale shown in the inset. (b) The LDOS 
at the chain end plotted against $\theta$ for $Q=1$
where $\tilde{\mu}=2.5\Delta$ and ${\bf \Gamma}$ is the same as above. Since there is now a
single Majorana fermion (MF) at each end the zero-bias peak does not split. Although the system is
no longer in class BDI, it is still a class-D topological
superconductor with zero or one MF at each end. We use arbitrary units for the LDOS in these plots. \label{fig:ldosBy}} 
\end{figure}

While the above results demonstrate that it is not necessary to fine-tune
the chemical potential or Zeeman spin-splitting to generate a
zero-energy peak in LDOS (and consequently a ZBP in STM measurements)
at the 
ends of the FM chain, a component of ${\bf \Gamma}$ perpendicular to
the $x$-$z$ plane breaks the chiral symmetry. To assess the effects of
misalignment of the Zeeman splitting 
(which can, for example, be generated by a suitably applied external
magnetic field), 
we plot in Fig.~\ref{fig:ldosBy} the LDOS against $\theta$ where
now we choose ${\bf \Gamma}=3\Delta\left(\sin\theta{\bf
  e}_y+\cos\theta{\bf e}_z\right)$.  
The zero-energy LDOS peak at the end of the chain splits into two peaks at finite energy by a non-zero $\theta$ only in the phase $Q=2$.
As $\theta$ is tuned up, the magnitude of the splitting
first increases, then decreases, and finally vanishes with a concomitant
disappearance of the localized peak. This can be understood from the
observation that the $y$-component of ${\bf \Gamma}$ has an additional
effect of suppressing 
the spectral gap of the system and since the splitting is bounded
by the size of the spectral gap, the size of the splitting can never
reach a large value. Therefore the splitting of the zero-energy LDOS
peak due to a misalignment of the Zeeman term is always small. No such
splitting should be observable in the phase with 
$Q=1$. This is because in these regions of the phase diagram the ends
of the chain host a single MF at each end, and thus the ZBP persists. 
Although the system is no longer in class BDI, it reduces to a class-D
topological superconductor with zero or one MF at each end.

Before concluding, we comment on the connection to previous works. Our
platform bears a superficial resemblance to proposals where the
impurity band formed by
a chain of magnetic impurities deposited on
the surface of  an $s$-wave superconductor naturally resides in a
topological phase 
\cite{Choy2011Majorana,Nadj-Perge2013Proposal}, due to the self-tuned
formation of a spin helix resulting from the RKKY interaction mediated
by the quasiparticles in the superconductor
\cite{Klinovaja2013Topological,Pientka2013Topological,Pientka2014Unconventional,Kim2014Helical,Vazifeh2013Self-Organized}. Our
system nevertheless differs from this class of proposals in
several fundamental ways: (i) due to large direct hopping
between the atoms in the chain, we are focused on the electronic
states on the chain itself, and not on the impurity band in the
superconductor; (ii) we assume a large direct exchange 
interaction between the impurities such that 
they are in a ferromagnetic arrangement and RKKY interactions play no
role; and (iii) the topological state in our system follows from
spin-orbit coupling in the superconductor which induces 
a triplet pairing term in the chain. Our work has much
closer connection to previous proposals
\cite{Chung2011Topological,Duckheim2011Andreev} in which half-metals
are proximity-coupled to spin-orbit coupled superconductor
surface. These can be considered as the special case of our device
where only one band of the ferromagnetic chain is
occupied and a single MF is present at each end of the
chain. Our crucial improvement over these schemes is that we
have demonstrated that a different topological phase with two MFs at
each end can be realized when both bands of the chain are occupied, which is
certainly much less restrictive than requiring a half-metal chain. We
also note the relation to semiconductor-superconductor 
heterostructure proposals 
\cite{Lutchyn2010Majorana,Oreg2010Helical} where the spin-orbit
coupling is in the Majorana wire itself and the spin-splitting is
produced by an explicit external magnetic field in sharp contrast to
our system.  There has actually been one published experiment
\cite{Wang2010Interplay} (and a related theoretical analysis
\cite{Takei2012Microscopic}) involving transport studies on a somewhat
related system with a Co nanowire sandwiched between superconducting
electrodes although the specific MF issues of interest in the current
paper were not investigated in these works. A possible experimental
verification of our predictions has recently been reported in
measurements of robust ZBPs at the end of ferromagnetic Fe chain on
superconducting Pb \cite{NadjPerge2014Science}. While very suggestive,
the actual relation to our proposal is unclear [see Ref.~\cite{Brydon2015Topological}] and requires detailed
modelling which is well beyond the scope of the current work.

In conclusion, we consider a FM chain deposited on the surface of a
bulk $s$-wave superconductor with strong spin-orbit coupling. We
establish the generic existence of a zero-energy peak in the LDOS at
the ends of the chain in this system. The zero-energy peak in the LDOS
should 
be accessible in STM experiments which should reveal a pronounced ZBP
from the chain ends but not from the regions away from the ends. We
show that the ZBP is due to the existence of one (odd) or two (even)
MFs localized at the same end protected by a topological chiral
symmetry. In this picture an STM experiment on the ends of a FM chain
deposited on the surface a bulk superconductor (with strong spin-orbit
coupling) will almost always show a pronounced ZBP, indicating the
existence of one or two MFs at each end depending on the relative
magnitudes of the ferromagnetic moment and the chemical potential. 

\npsection{Methods}

{\footnotesize 
Here we present a derivation of the effective
chain Hamiltonian. Starting from the general expression for the
pseudospin states, we derive the Green's function in the orbital-spin
basis. We then use this to evaluate the lowest-order self-energy
correction to the chain states due to the proximity effect. Evaluated
in the static limit, we add the self-energy to the bare chain
Hamiltonian to obtain the effective model studied in the main text.

\npsection{Pseudospin basis.}
The pseudospin state is expressed in terms of the orbital-spin states as
\begin{equation}
|{\bf k},\varsigma\rangle = \sum_{\sigma=\uparrow,\downarrow}\left\{B^s_{\varsigma,\sigma}({\bf k})|s,{\bf
  k},\sigma\rangle + B^p_{\varsigma,\sigma}({\bf k})|p,{\bf
  k},\sigma\rangle\right\}\,. \label{eq:supppseudospin}
\end{equation}
The pseudospin index $\varsigma=\pm$ transforms as a spin under
time-reversal (${\cal T}$) and inversion (${\cal I}$) symmetries. From
\begin{gather}
{\cal T}|s,{\bf 
  k},\sigma \rangle = \sigma |s,-{\bf k},-\sigma\rangle\,,
\qquad {\cal I}|s,{\bf k},\sigma\rangle = |s,-{\bf
  k},\sigma\rangle\,, \\
{\cal T}|p,{\bf 
  k},\sigma \rangle = \sigma |p,-{\bf k},-\sigma\rangle\,,
\qquad {\cal I}|p,{\bf k},\sigma\rangle = -|p,-{\bf
  k},\sigma\rangle\,,
\end{gather}
we deduce the relations obeyed by the coefficients
\beqarray
\text{inversion:} \quad \Rightarrow \quad B^s_{\varsigma,\sigma}({\bf k}) &=&
B^{s}_{\varsigma,\sigma}(-{\bf k})\,,\nonumber\\
 \quad B^p_{\varsigma,\sigma}({\bf k}) &=&-B^{p}_{\varsigma,\sigma}(-{\bf k})\,, \label{eq:suppBinv}\\
\text{time-reversal:} \quad \Rightarrow \quad  B^s_{\varsigma,\sigma}({\bf k}) &=&
\varsigma\sigma[B^{s}_{-\varsigma,\overline{\sigma}}(-{\bf k})]^\ast\,,\nonumber\\
 \quad B^p_{\varsigma,\sigma}({\bf k}) &=&\varsigma\sigma[B^{p}_{-\varsigma,\overline{\sigma}}(-{\bf
    k})]^\ast\,. \label{eq:suppBtri}
\eeqarray
Expressed as a matrix, we have the general forms of $\hat{B}^s({\bf
  k})$ and $\hat{B}^p({\bf k})$
\beqarray
\hat{B}^s({\bf k}) & = & \alpha^s_{\bf k} + i\pmb{\beta}^s_{\bf
  k}\cdot\hat{\pmb{\sigma}}\,, \label{eq:suppBscoeff} \\
\hat{B}^p({\bf k}) & = & \alpha^p_{\bf k}\hat{\pmb{\sigma}}\cdot{\bf e}_{\bf k} + \pmb{\beta}^p_{\bf
  k}\cdot\left(\hat{\pmb{\sigma}}\times{\bf e}_{\bf k}\right) +
i\pmb{\gamma}^p_{\bf k}\cdot{\bf e}_{\bf k}\,, \label{eq:suppBpcoeff}
\eeqarray
where ${\bf e}_{\bf k}$ is the unit vector in the direction of ${\bf
  k}$, and the coefficients $\alpha^s_{\bf k}$, $\pmb{\beta}^s_{\bf k}$,
$\alpha^p_{\bf k}$, $\pmb{\beta}^p_{\bf k}$, and $\pmb{\gamma}^p_{\bf
  k}$ are real and even functions of ${\bf k}$.

We can further constrain the forms of  $\hat{B}^s({\bf
  k})$ and $\hat{B}^p({\bf k})$ by considering mirror symmetries. We
  assume that the crystal has mirror planes perpendicular to the $x$,
  $y$, and $z$ axes. We assume that the pseudospin transforms like a 
  spin under mirror reflection, i.e.
\beq
{\cal M}_{\mu}|{\bf k},\sigma\rangle = \hat{\sigma}^\mu|{\bf
  -k},\sigma\rangle\,. 
\eeq
We assume that this also hold for the orbital states, except that the
odd-parity orbital is odd under mirror reflection in the plane
perpendicular to the $z$ axis. It can then be shown that
\beqarray
\hat{B}^s({\bf k}) & = & a^{s}_{{\bf k}}\hat{\sigma}^0 +
i\mbox{sgn}(k_yk_z)b^{s}_{{\bf k}}\hat{\sigma}^x\nonumber\\
&& +i\mbox{sgn}(k_xk_z)c^s_{{\bf k}}\hat{\sigma}^y +
i\mbox{sgn}(k_xk_y)d^s_{{\bf k}}\hat{\sigma}^z \label{eq:mirrorBscoeff} \\
\hat{B}^p({\bf k}) & = & \mbox{sgn}(k_z)\left[ia^p_{{\bf
    k}}\hat{\sigma}^0 + \mbox{sgn}(k_yk_z)b^p_{{\bf
    k}}\hat{\sigma}^x\right.\nonumber\\
&& \left.+ \mbox{sgn}(k_xk_z)c^p_{{\bf
    k}}\hat{\sigma}^y + \mbox{sgn}(k_xk_y)d^p_{z,{\bf
    k}}\hat{\sigma}^z\right] \label{eq:mirrorBpcoeff}
\eeqarray
where the coefficients $a^{s}_{\bf k}$, {\it etc.} are real functions and
even under mirror reflection.

\npsection{Green's function in orbital-spin basis.}
Expressed in the basis $\Psi_{\text{pseudo}}({\bf k}) = (c_{{\bf
    k},+},c_{{\bf k},-},c^{\dagger}_{{\bf 
    k},-},-c^\dagger_{{\bf k},+})^T$, where $c_{{\bf
    k},\varsigma}$ is the annihilation operator for the state with
momentum ${\bf k}$ and pseudospin $\varsigma$, the pseudospin Green's
function of the bulk superconductor is the $4\times4$ matrix
\beq
G_{\text{pseudo}}({\bf k},\omega) = \frac{\omega\hat{\tau}_0 + \xi_{\bf
    k}\hat{\tau}_z + \Delta_0\hat{\tau}_x}{\omega^2 - \xi_{\bf k}^2
  - \Delta_0^2}
\eeq
where $\xi_{\bf k}$ is the normal state dispersion, $\Delta_0$ is the
superconducting gap, and $\hat{\tau}_{\mu}$ are the Pauli matrices in
Nambu space. 
From equation (\ref{eq:supppseudospin}) we have the relation
\beq
\Psi_{\text{pseudo}}({\bf k}) = \left(\begin{array}{cc}\check{B}^\ast_{\bf k} & 0 \\
0 & \check{B}^\ast_{\bf k}\end{array}\right)\Psi_{\text{orb}}({\bf k}) \label{eq:supppseudotospinorb}
\eeq
where
\beq
\Psi_{\text{orb}}({\bf k}) = \left(s_{{\bf
      k},\uparrow}\,,  s_{{\bf k},\downarrow} \,,  p_{{\bf
      k},\uparrow}\,, p_{{\bf k},\downarrow} \,,  s^\dagger_{-{\bf
      k},\downarrow}\,,  -s^\dagger_{-{\bf k},\uparrow} \,,  p^\dagger_{-{\bf
      k},\downarrow}\,,  -p^\dagger_{-{\bf k},\uparrow}\right)^T
\eeq
is the spinor of creation and annihilation operators in the
orbital-spin basis, where $s_{{\bf k},\sigma}$ ($p_{{\bf k},\sigma}$)
destroys an electron with momentum ${\bf k}$ and spin $\sigma$ in the
$s$ ($p$) orbital, and 
\beq
\check{B}_{\bf k} = \left(\begin{array}{cc}\hat{B}^s({\bf k}) & \hat{B}^p({\bf k})\end{array}\right)
\eeq
is a $2\times4$ matrix, with $\hat{B}^s({\bf
  k})$ and $\hat{B}^p({\bf k})$ as defined above. 
Using equation (\ref{eq:supppseudotospinorb}), we express the Green's function
in the orbital basis  as
\beq
G_{\text{orb}}({\bf k},\omega) = \left(\begin{array}{cc}\check{B}^T_{\bf k} & 0 \\
0 & \check{B}^T_{\bf k}\end{array}\right)G_{\text{pseudo}}({\bf
  k},\omega)\left(\begin{array}{cc}\check{B}^\ast_{\bf k} & 0 \\
0 & \check{B}^\ast_{\bf k}\end{array}\right)\,, \label{eq:suppGorb}
\eeq
where $G_{\text{orb}}({\bf k},\omega)$ is an $8\times8$ matrix. It is
important to note that since this Green's function is obtained from
the pseudospin Green's function $G_{\text{pseudo}}({\bf k},\omega)$,
it is only valid close to the Fermi energy. The full orbital-spin
Green's function contains terms from the additional band
composed from the $s$ and $p$ orbitals, but since this band is assumed to
lie far away from the Fermi surface we ignore them.

\npsection{Proximity effect.}
The proximity effect in the chain due to the tunneling into the
superconductor is accounted for by the self-energy
\beq
\Sigma(x,x';\omega) = {\bf T}G_{\text{orb}}(x,x';\omega){\bf T}^\dagger
\eeq
where the $4\times8$ matrix ${\bf T}$ describes the tunneling between the
orbital-spin states of the superconductor and the ferromagnetic chain
\beq
{\bf T} = \left(\begin{array}{cccc}
t_s\hat{1} & t_p \hat{1} & 0 & 0 \\
0 & 0 & -t_s\hat{1} & -t_p \hat{1}\end{array}\right)\,.
\eeq
For simplicity, we approximate the Green's function of the
superconductor at the surface by the bulk Green's
function equation (\ref{eq:suppGorb}). This is a reasonable approximation for
the conventional superconductors considered here. We hence obtain
\beqarray
\Sigma(x,x';\omega) &=& \int\frac{d^3k}{(2\pi)^3}{\bf
  T}G_{\text{orb}}({\bf k},\omega){\bf T}^\dagger e^{ik_x(x-x')}\nonumber\\
 &=& \int\frac{d^3k}{(2\pi)^3}\Sigma({\bf k},\omega) e^{ik_x(x-x')} 
\eeqarray
After straightforward manipulation, we find
\beq
\Sigma({\bf k},\omega) = \frac{1}{\omega^2 - \xi_{\bf k}^2 -
  \Delta_0^2}\left(\begin{array}{cc} 
(\omega + \xi_{\bf k})\hat{\Xi}({\bf k})
  & -\Delta_0 \hat{\Xi}({\bf k}) \\
-\Delta_0 \hat{\Xi}({\bf k})
  & (\omega - \xi_{\bf k})\hat{\Xi}({\bf k}) \end{array}\right)
\eeq
where
\beqarray
\hat{\Xi}({\bf k}) &=& t^2_s[\hat{B}^s({\bf k})]^T[\hat{B}^s({\bf k})]^\ast
 + t^2_p[\hat{B}^p({\bf k})]^T[\hat{B}^p({\bf k})]^\ast\nonumber\\
 &&+ t_st_p[\hat{B}^s({\bf k})]^T[\hat{B}^p({\bf k})]^\ast
 + t_st_p[\hat{B}^p({\bf k})]^T[\hat{B}^s({\bf k})]^\ast\,.
\eeqarray
The first two terms in $\hat{\Xi}({\bf k})$ describe tunneling
processes involving only one of the orbitals in the
superconductor. Using the properties~Eqs.(\ref{eq:suppBinv})
and~(\ref{eq:suppBtri}), the matrix products here can be shown to be
proportional to the unit matrix, and to be even functions of ${\bf k}$.
The last two terms, on the other hand, arise from tunneling processes
involving both orbitals, where the matrix products are proportional to
the Pauli matrices and are odd in ${\bf k}$. In particular, this
introduces spin-triplet pairing correlations into the ferromagnetic
chain. 

\npsection{Effective Hamiltonian.}
To obtain an effective Hamiltonian for the ferromagnetic chain
including the proximity effect, we add the self-energy term
evaluated at $\omega = 0$ to the bare chain Hamiltonian, i.e.
\beq
H^{\text{eff}}_{\text{chain}}(x,x') = H^{(0)}_{\text{chain}}(x,x') + \Sigma(x,x';\omega=0) \label{eq:suppHeff}
\eeq
where the original chain Hamiltonian is written
\beq
H^{(0)}_{\text{chain}}(x,x') = t(x,x')\hat{\tau}_z + \pmb{\Gamma}\cdot\hat{\pmb{\sigma}}\delta_{x,x'}\,.
\eeq
Here $t(x,x') = -\mu\delta_{x,x'} - t(\delta_{x,x'+1} +
\delta_{x,x'-1})$ describes the normal state dispersion, while
$\pmb{\Gamma}\cdot\hat{\pmb{\sigma}}$ is the Zeeman splitting due to the
ferromagnetism. The proximity effect renormalizes the dispersion, and
also introduces an antisymmetric spin-orbit coupling, and spin-singlet
and spin-triplet pairing potentials,
\beqarray
\Sigma(x,x';\omega=0) &=& \delta t(x,x')\hat{\tau}_z + {\bf
  g}(x,x')\cdot\hat{\pmb{\sigma}}\hat{\tau}_z \nonumber\\
  &&+\Delta^s(x,x')\hat{\tau}_x + {\bf d}(x,x')\cdot\hat{\pmb{\sigma}}\hat{\tau}_x\,,
\eeqarray
where we have the general expressions
\begin{widetext}\beqarray
\delta t(x,x') & = &
-\int\frac{d^3k}{(2\pi)^3}e^{ik_x(x-x')}\frac{\xi_{\bf k}}{\xi_{\bf k}^2 + \Delta_0^2}\left\{t^2_s[\hat{B}^s({\bf k})]^T
[\hat{B}^s({\bf k})]^\ast + t^2_p[\hat{B}^p({\bf k})]^T
[\hat{B}^p({\bf k})]^\ast\right\}\,,\\
{\bf g}(x,x')\cdot\hat{\pmb{\sigma}} & = &
-\int\frac{d^3k}{(2\pi)^3}e^{ik_x(x-x')}\frac{\xi_{\bf k}}{\xi_{\bf k}^2 + \Delta_0^2}\left\{t_st_p[\hat{B}^s({\bf k})]^T
[\hat{B}^p({\bf k})]^\ast + t_st_p[\hat{B}^p({\bf k})]^T
[\hat{B}^s({\bf k})]^\ast\right\} \,,\\
\Delta^s(x,x') &=&
\int\frac{d^3k}{(2\pi)^3}e^{ik_x(x-x')}\frac{\Delta_0}{\xi_{\bf k}^2 + \Delta_0^2}\left\{t^2_s[\hat{B}^s({\bf k})]^T
[\hat{B}^s({\bf k})]^\ast + t^2_p[\hat{B}^p({\bf k})]^T
[\hat{B}^p({\bf k})]^\ast\right\} \,,\\
{\bf d}(x,x')\cdot\hat{\pmb{\sigma}} &=&
\int\frac{d^3k}{(2\pi)^3}e^{ik_x(x-x')}\frac{\Delta_0}{\xi_{\bf k}^2 + \Delta_0^2}\left\{t_st_p[\hat{B}^s({\bf k})]^T
[\hat{B}^p({\bf k})]^\ast + t_st_p[\hat{B}^p({\bf k})]^T
[\hat{B}^s({\bf k})]^\ast\right\}\,.
\eeqarray\end{widetext}
Although these expressions are quite complicated, we can nevertheless
make some generic observations. Firstly, we note that the
renormalization of the dispersion and the singlet pairing potential
arise only from the intra-orbital tunneling processes. On the other
hand, the inter-orbital processes are responsible for the spin-orbit
coupling and the triplet gap. The opposite parity of the $s$ and $p$
orbitals is crucial in obtaining these terms; tunneling into
orbitals of the same parity could only give even-parity contributions
to the self-energy. Furthermore, we observe that the induced
spin-orbit coupling vector is always parallel to the triplet ${\bf d}$
vector, i.e. ${\bf g}(x,x')||{\bf d}(x,x')$. We
expect that the 
pairing terms are generally much larger than the normal-state
corrections, however, due to the factor of $\xi_{\bf k}$ in the
integrals of the latter. We henceforth ignore $\delta t(x,x')$ and
${\bf g}(x,x')$ in constructing the effective Hamiltonian.

To derive a tractable model for the chain, we first assume that the
proximity-effect corrections $\Sigma(x,x';\omega=0)$ are negligible
for $x$ and $x'$ further apart than nearest neighbors. We then choose
coefficient
matrices $\hat{B}^s({\bf k})$ and $\hat{B}^p({\bf k})$. For simplicity
we take  $\alpha^s_{\bf k} = \alpha^s $, $\pmb{\beta}^p_{\bf k} =
\beta^p{\bf e}_z$, and all other coefficients vanishing in
Eqs.~(\ref{eq:suppBscoeff}) and (\ref{eq:suppBpcoeff}); this is
equivalent to $a^s_{\bf k} = \alpha^s$, $b^p_{\bf k} = |k_y|\beta^p$,
$c^p_{\bf k} = -|k_x|\beta^p$ in Eqs.~(\ref{eq:mirrorBscoeff}) and
(\ref{eq:mirrorBpcoeff}). Other choices of coefficients can only change the orientation of ${\bf d}(x,x')$. We then find
\begin{multline}
t^2_s[\hat{B}^s({\bf k})]^T
[\hat{B}^s({\bf k})]^\ast + t^2_p[\hat{B}^p({\bf k})]^T
[\hat{B}^p({\bf k})]^\ast \\ =  t_s^2(\alpha^s)^2 +
t_p^2(\beta^p)^2(\hat{k}_x^2 + \hat{k}_y^2) \,,\end{multline}
\begin{multline}
t_st_p[\hat{B}^s({\bf k})]^T
[\hat{B}^p({\bf k})]^\ast + t_st_p[\hat{B}^p({\bf k})]^T
[\hat{B}^s({\bf k})]^\ast  \\=  2t_st_p\alpha^s\beta^p(\hat{\sigma}_x\hat{k}_y + \hat{\sigma}_y\hat{k}_x)\,,
\end{multline}
where $\hat{k}_\nu = k_{\nu}/|{\bf k}|$.
We hence obtain the gap functions
\beqarray
\Delta^s(x,x') & = & \Delta \delta_{x,x'} +
\frac{1}{2}\tilde{\Delta}\left(\delta_{x,x'+1} +
\delta_{x,x'-1}\right) \,, \label{eq:suppsingletgap} \\
{\bf d}(x,x')\cdot\pmb{\hat{\sigma}} & = &
\frac{1}{2}\tilde{\Delta}^{(t)}\left\{\delta_{x,x'-1} -
\delta_{x,x'+1}\right\}\hat{\sigma}_y\,, \label{eq:supptripletgap}
\eeqarray
where
\beqarray
\Delta & = & \int\frac{d^3k}{(2\pi)^3}\frac{\Delta_0}{\xi_{\bf k}^2 +
  \Delta_0^2}\left\{t_s^2(\alpha^s)^2 + t_p^2(\beta^p)^2(\hat{k}_x^2 + \hat{k}_y^2)\right\} \,,\\
\tilde{\Delta} & = & \int\frac{d^3k}{(2\pi)^3}e^{ik_xa}\frac{\Delta_0}{\xi_{\bf k}^2 +
  \Delta_0^2}\left\{t_s^2(\alpha^s)^2 + t_p^2(\beta^p)^2(\hat{k}_x^2 + \hat{k}_y^2)\right\} \,,\\
\tilde{\Delta}^{(t)} & = & \int\frac{d^3k}{(2\pi)^3}e^{ik_xa}\frac{\Delta_0}{\xi_{\bf k}^2 +
  \Delta_0^2}\left\{2t_st_p\alpha^s\beta^p \hat{k}_x\right\}\,.
\eeqarray
Here $a$ is the lattice spacing of the chain.

Finally, we insert the pairing potentials equations (\ref{eq:suppsingletgap})
and~(\ref{eq:supptripletgap}) into equation (\ref{eq:suppHeff}) and
transform to momentum space to obtain the effective Hamiltonian which is
studied in the main text
\beqarray
H^{\text{eff}}_{\text{chain}}(k_x) &=& \left(-2t\cos
k_{x}-\mu\right)\hat{\tau}_{z}+{\bf \Gamma}\cdot\hat{\pmb{\sigma}}\nonumber\\
&&+\left(\Delta+\tilde{\Delta}\cos k_{x}\right)\hat{\tau}_{x}
+\tilde{\Delta}^{(t)}\sin k_{x}\hat{\sigma}_{y}\hat{\tau}_{x}\,. 
\eeqarray
}

\npsection{Acknowledgements} 
  This work is supported by NSF-JQI-PFC, LPS-CMTC, and Microsoft Q. Funding for Open Access provided by the UMD Libraries Open Access Publishing Fund.

\npsection{Author contributions} 
  S.D.S., J.D.S. and S.T. contributed to the conceptual developments. S.T., P.M.R.B. and H.Y.H. wrote the mainmanuscript text and prepared the figures. All authors reviewed and edited the manuscript.

\npsection{Additional information} 
 Competing financial interests: The authors declare no competing financial interests.
\clearpage
\bibliography{paper}
\bibliographystyle{naturemag}

\end{document}